\documentstyle[prb, aps]{revtex}

\begin{document}

 \draft

\title{Premartensitic Transition in 
${\bf Ni}_{2+x}{\bf Mn}_{1-x}{\bf Ga}$ 
Heusler Alloys}

 \author {V. V. Khovailo$^1$ \and T. Takagi$^1$ \and A. D. Bozhko$^2$
         \and M. Matsumoto$^3$ \and J. Tani$^1$ \and V. G. Shavrov$^4$}
 \address {$^1$Institute of Fluid Science, Tohoku University, Sendai 980--77, Japan}
  
 \address{$^2$Physics Faculty, Moscow State University, Moscow 119899, Russia}
 
 \address {$^3$Institute for Advanced Materials Processing, Tohoku University, 
 Sendai 980--77, Japan}

 \address {$^4$Institute of 
 Radioengineering and Electronics of RAS, Moscow 103907, Russia}

 \maketitle

 \begin {abstract}
The temperature dependencies of the resistivity and magnetization of 
a series of ${\rm Ni}_{2+x}{\rm Mn}_{1-x}{\rm Ga}$ 
({\it x} = 0 - 0.09) alloys were investigated. Along with the 
anomalies associated with ferromagnetic and martensitic transitions, 
well-defined anomalies were observed at the temperature of premartensitic 
transformation. The premartensitic phase existing in a temperature 
range 200 -- 260 K in the stoichiometric ${\rm Ni}_2{\rm MnGa}$ is 
progressively suppressed by the martensitic phase with increasing Ni content and 
vanishes in ${\rm Ni}_{2.09}{\rm Mn}_{0.91}{\rm Ga}$ composition.
\end{abstract}


In ${\rm Ni}_2{\rm MnGa}$, like in many other Heusler alloys containing 
manganese, the indirect exchange interaction between magnetic ions results 
in ferromagnetism which is usually described in terms of the local magnetic 
moment at the Mn site \cite{Plog}. For the stoichiometric ${\rm Ni}_2{\rm MnGa}$ 
a structural transition of the martensitic type from the parent cubic to 
a complex tetragonally based structure occurs at $T_M = 202$ K while 
ferromagnetic ordering sets at $T_C = 376$ K \cite{Web}. The martensitic 
transition temperature $T_M$ was found to be sensitive to the composition, 
and values of $T_M$ between 175 and 450 K have been reported. A specific 
feature of the Ni-Mn-Ga system is that in alloys with a high $T_M$ ($> 270$ K) 
a number of intermartensite transformations can be induced by an external 
stress \cite{Mart,ANV}, whereas in the alloys with lower $T_M$ the 
martensitic transition is preceded by a weakly first order premartensitic 
phase transition \cite{Planes,Sten,AGon,Cast}. The inelastic neutron scattering 
experiments \cite{Zhel} performed on stoichiometric 
${\rm Ni}_2{\rm MnGa}$ showed the existence of 
a soft $[\xi\xi0]{\rm TA}_2$ phonon mode in a wide temperature interval, 
which is common to martensitic alloys having bcc structure.  An important 
observation in these measurements was that the ${\rm TA}_2$ phonon branch 
at a wave vector of $\xi_0 \approx 0.33$ incompletely condenses at the 
premartensitic transition $T_P \approx 260$ K, well above the martensitic 
transition $T_M = 220$ K. On cooling from $T_P$ to $T_M$ the frequency of 
the soft mode increases. This would suggest that the soft mode formation is 
associated with the premartensitic phase transformation rather then with the 
martensitic one. By transmission electron microscopy observation \cite{Cesar} 
it was established that the premartensitic phase consists in a 
micromodulated "tweed" structure without macroscopic tetragonal distortions 
so that the parent cubic symmetry is preserved. The modulation of the 
premartensitic phase was found to correspond to the wave vector 
$\xi_0 \approx 0.33$.

It is necessary to stress that although no premartensitic transition was found 
in compositions with high $T_M$, the precursor phenomenon (softening of 
the $[\xi \xi 0]{\rm TA}_2$ phonon branch at the wave vector 
$\xi_0 \approx 0.33$) has been clearly observed by inelastic neutron 
scattering \cite{Stu15,Stu16}. Except for the partial condensation of 
the ${\rm TA}_2$ phonon branch observed by Zheludev {\it et al}.,  
the only essential difference between the inelastic neutron scattering results 
performed on three samples of different 
stoichiometry \cite{Zhel,Stu15,Stu16} is that the width in $\xi$ where 
softening occurs becomes broader as $T_M$ increases.

At the presence the reason why the premartensitic transformation is observed 
only in the alloys with $T_M \le 260$ K seems to be unclear. There is strong 
evidence \cite{Planes,AGon,Stu15} that the magneto-elastic interaction 
plays a crucial role in the formation of the premartensitic phase. Also, 
the theoretical investigation \cite{Cast} predicts that the premartensitic 
transition is observed only in the case of large values of the magneto-elastic 
coupling parameter. From the results of magnetic and transport measurements 
of several near-stoichiometric ${\rm Ni}_{2-x}{\rm Mn}_{1+x}{\rm Ga}$ 
samples \cite{Zuo} it can be concluded that the premartensitic transition is 
less sensitive to the change of composition than the martensitic one. 
This gives ground to suggest that a deviation from the stoichiometry can lead to 
the disappearance of the premartensitic transition in some critical composition.

The importance of the conduction electron density in stabilizing the 
Heusler structure was noted a long ago and, in particular, it was 
suggested \cite{Smit} that the structure is stabilized because the 
Fermi surface touches the Brillouin zone boundary.
From this point of view a partial substitution of Mn for Ni 
in ${\rm Ni}_{2+x}{\rm Mn}_{1-x}{\rm Ga}$ alloys 
resulting in an increase of the conduction 
electron density should increase the martensitic transition temperature.
This makes it possible to determine the compositional dependencies of $T_M$ and $T_P$. 
 For this purpose 
we have studied temperature dependencies of electrical resistivity and 
magnetization in ${\rm Ni}_{2+x}{\rm Mn}_{1-x}{\rm Ga}$ alloys in 
which Mn is partially substituted for Ni in the range of $x = 0 - 0.09$.

The polycrystalline ${\rm Ni}_{2+x}{\rm Mn}_{1-x}{\rm Ga}$ ingots were 
prepared by a conventional arc-melting method in Ar atmosphere. The ingots 
were homogenized in vacuum at 1100 K for 9 days and slowly cooled to room 
temperature. Room temperature X-ray diffraction showed single-phase cubic 
structure of the alloys. The samples for the resistivity and magnetization 
measurements were spark cut from the ingots and were 
of $0.6\times1\times6$ ${\rm mm}^3$ dimensions. The electrical resistivity was 
measured in a temperature range 100 -- 450 K by ac four-terminal method. 
Magnetic measurements were done by a SQUID magnetometer.

The temperature dependencies of resistivity $\rho$ measured at cooling 
in ${\rm Ni}_{2+x}{\rm Mn}_{1-x}{\rm Ga}$ ($x = 0 - 0.09$) samples 
are shown in Fig. 1. Except for the ${\rm Ni}_{2.09}{\rm Mn}_{0.91}{\rm Ga}$ 
composition, the $\rho (T)$ dependencies in the other samples exhibit three 
well-defined anomalies at temperatures corresponding 
to the ferromagnetic ($T_C$), premartensitic ($T_P$) 
and martensitic ($T_M$) transitions. With deviation from the stoichiometry the change 
in a slope of the resistivity curve related to the ferromagnetic transition 
shifts to lower temperatures, and the jump-like behavior of $\rho$ related to 
the martensitic transformation shifts to higher temperatures. This decrease 
of $T_C$ and increase of $T_M$ with increasing Ni content is in good agreement 
with \cite{ANV18}. While the change in a slope at $T_C$ and jump-like behavior 
at $T_M$ are typical features of the resistivity in Ni-Mn-Ga alloys, the 
prominent anomaly at $T_P$ has been observed for the first time. Namely, 
the resistivity demonstrates either a well-defined peak 
(${\it x} = 0 - 0.04$ samples) or a two-step-like behavior 
(${\it x} = 0.06 - 0.08$ samples) at this temperature. The premartensitic 
transition temperature $T_P$ was determined either as the peak or as the 
bend of two-step-like resistivity curve. Instead of $T_C$ and $T_M$, the 
premartensitic transition temperature $T_P$ does not shift with Ni excess 
and $T_P \approx 260$ K remains essentially the same for all alloys in the 
studied range of compositions (Fig. 2). 

In the compositions with low 
Ni content $T_M$ and $T_P$ are well separated and the premartensitic 
transition manifests itself as a well-defined peak on the resistivity 
curve. Increasing Ni content results in enhancement of $T_M$ and, as a 
consequence, only the right (high-temperature) part of the peak is 
presented on $\rho (T)$ for the ${\it x} = 0.06 - 0.08$ samples (Fig. 1). 
The left (low-temperature) part of the peak can be not detectable in 
the martensitic phase because all physical characteristics of the 
materials (crystal structure, Fermi surface, mean free path and so on) 
drastically change upon the martensitic transformation.  
It is also interesting to compare the behavior of the resistivity at $T_M$ 
and $T_P$ measured at heating and cooling 
for ${\rm Ni}_{2.04}{\rm Mn}_{0.96}{\rm Ga}$ 
and ${\rm Ni}_{2.08}{\rm Mn}_{0.92}{\rm Ga}$ (Fig. 3). In the case 
of ${\rm Ni}_{2.04}{\rm Mn}_{0.96}{\rm Ga}$, the resistivity shows a 
significant temperature hysteresis at $T_M$ whereas the temperature 
hysteresis at $T_P$ is small, which is in agreement with published 
results \cite{Planes,AGon,Zhel}. In 
the ${\rm Ni}_{2.08}{\rm Mn}_{0.92}{\rm Ga}$ sample the two-step-like 
anomaly observed at cooling completely disappears upon warming up and 
the resistivity smoothly decreases down to the extreme value. Moreover, 
the hysteretic feature seen in the upper part gradually narrows and 
vanishes in the down part. This testifies that the observed two-step-like 
behavior of the resistivity is not a peculiar feature of the martensitic 
transition in the $ x = 0.06 - 0.08$ samples but is the outcome of 
simultaneously occurring premartensitic and martensitic transitions.
The anomaly of resistivity corresponding 
to $T_P$ had been completely taken up by the martensitic phase for the 
highest Ni composition in the series, ${\rm Ni}_{2.09}{\rm Mn}_{0.91}{\rm Ga}$, 
and resistivity shows only two anomalies corresponding to the ferromagnetic 
and martensitic phase transitions (Fig. 1). In our opinion, these 
findings are consistent with the absence of the 
premartensitic transition in Ni-Mn-Ga alloys with a high $T_M$ 
temperature \cite{Stu15,Stu16}.

The temperature dependencies of magnetization $M$ measured in a 100 Oe magnetic 
field also showed clear anomalies at $T_P$ in the $x = 0 - 0.04$ samples. 
However, such anomalies were not observed, neither upon cooling nor upon 
heating, in the $x = 0.06 - 0.08$ samples. 
Actually, the absence of them upon warming up agrees with 
the corresponding resistivity data. 
Since in the $x = 0.06 - 0.08$ samples the martensitic and premartensitic 
transitions are merged, the related to $T_P$ anomalies of $M$ are not 
observed upon cooling probably because they are masked by the 
drastic decrease of $M$ occuring upon the martensitic transformation.
An example of $M(T)$ dependence for the compositions with 
separated $T_P$ and $T_M$ is shown in Fig. 4.
 $M(T)$ revealed a pronounced 
dip in magnetization of the sample at $T_P$. The drastic decrease in 
magnetization at $\approx 227$ K, shown in the inset of Fig. 4, is common for 
transformations from ferromagnetic martensite to ferromagnetic austenite 
in Ni-Mn-Ga alloys and will be not discussed here. The temperature of the 
premartensitic transition $T_P$ is equal to 262 K, which is in excellent 
agreement with the corresponding resistivity data. Instead of a smooth 
diminution of the magnetization at $T_M$, the premartensitic transition is 
attended by a complicated behavior of the magnetization. The rapid 
decrease of magnetization in the temperature interval from 268 to 262 K 
can be a consequence of the freezing of the atomic displacements 
related to the soft $\frac{1}{3}[110]$ phonon mode. In this case the appearance of 
 modulations of the cubic phase results in an increase of magnetic 
anisotropy which explains the drop of magnetization. The upturn in 
magnetization at temperatures below $T_P = 262$ K is due presumably to 
the ordering of the premartensitic phase. This suggestion is consistent 
with the evolution of the Bragg peak 
at $q = (\frac{1}{3} \frac{1}{3} 0)$ (Ref. 9) and the 
results of ultrasonic measurements \cite{Sten} which clearly demonstrated 
that the premartensitic phase becomes fully developed and ordered at 
temperatures below $T_P$.

The martensitic transformation in Ni-Mn-Ga alloys is believed to occur due to 
the contact between the Fermi surface and a Brillouin zone boundary \cite{Web}.  
In this sense the compositional dependence of $T_M$ 
in ${\rm Ni}_{2+x}{\rm Mn}_{1-x}{\rm Ga}$ can be understood as 
resulting from a change of the Fermi energy due to the increase in the conduction 
electrons concentration upon substitution of Mn for Ni. On other hand, it 
has been suggested \cite{Zhel} that the origin of the premartensitic 
transition lies in specific nesting features of the multiply connected Fermi 
surface. The effect of uniaxial stress \cite{Andre} on the anomalous phonon 
branch in ${\rm Ni}_2{\rm MnGa}$ results in the shifting of the phonon 
anomaly and the premartensitic transition temperature $T_P$ to 
higher {\it q}-values and temperatures. External stress can actually 
change the geometry of the Fermi surface and modifies the nesting 
vector.  The fact that $T_P$ does not depend 
on {\it x} in ${\rm Ni}_{2+x}{\rm Mn}_{1-x}{\rm Ga}$ alloys means 
presumably that the increase in the conduction electrons concentration does 
not affect significantly this particular part of the Fermi surface.

Summarizing our speculation about the compositional dependence of the 
martensitic and premartensitic phase transitions, we argue that the 
premartensitic phase is suppressed by the martensitic phase with 
deviation from the stoichiometry 
in ${\rm Ni}_{2+x}{\rm Mn}_{1-x}{\rm Ga}$ alloys. The premartensitic 
transition couples with the martensitic transition in 
the range of $x$ from 0.06 to 0.08 and completely vanishes at 264 K in the 
critical composition $x = 0.09$ 
of ${\rm Ni}_{2+x}{\rm Mn}_{1-x}{\rm Ga}$ alloys. However, we do not 
rule out the possibility that the premartensitic transition gives rise to 
an intermartensitic transition in alloys with higher Ni content. Actually, 
such a possibility has been mentioned in \cite{AGon}, where the authors 
analyze data collected from the literature for a broad range of Ni-Mn-Ga 
compositions. From this point of view it is also worth mentioning that 
in a simpler classification the experimentally observed modulations of 
the martensitic phase in Ni-Mn-Ga alloys correlate with the temperature 
of martensitic transition $T_M$ as follow \cite{Pons}: 5-layered martensite 
is observed in alloys with $T_M < 270$ K, whereas alloys with $T_M > 270$ K 
exhibit 7- or 10-layered martensitic structures. This characteristic martensitic 
transition temperature $T_M = 270$ K where the modulation of the martensitic 
phase changes from 5 to 7 (or 10) layers accords well with the critical 
temperature $T_P = 264$ K at which the premartensitic phase vanishes.

Very recently \cite{VDB} a phenomenological theory of structural and 
magnetic phase transitions in ${\rm Ni}_{2+x}{\rm Mn}_{1-x}{\rm Ga}$
alloys which takes into account strain, crystal lattice modulation, magnetic 
order parameter and interaction between these subsystems has been developed. 
It has been shown that the addition of an order parameter $\psi$ accounting for 
the crystal lattice modulation of the premartensitic phase to the general 
expression of the free energy \cite{ANV18} leads to an intermartensitic 
transition. On the compositional phase diagram this transition results 
from an extension of the line of the premartensitic transition into 
martensitic phase. The results of numerical calculations indicate that 
the premartensitic transition temperature as a function of concentration 
in ${\rm Ni}_{2+x}{\rm Mn}_{1-x}{\rm Ga}$ alloys intersects the 
line of the martensitic phase transition at $x \sim 0.11$. This agrees 
fairly well with the experimental value $x = 0.09$.

In conclusion, the most novel and significant findings of our experiments 
are that the premartensitic phase is progressively suppressed by 
the martensitic phase until it completely vanishes in the critical 
composition ${\rm Ni}_{2.09}{\rm Mn}_{0.91}{\rm Ga}$ at 264 K. If the suggested 
in \cite{Web} and \cite{Zhel} features of the Fermi surface 
indeed are responsible for $T_M$ and $T_P$, this means that the increase 
in the conduction electrons density does not modify the nesting 
vector and affects mainly the Fermi energy.

We are grateful to Professor A. N. Vasil'ev and Professor V. D. Buchel'nikov 
for helpful discussions. This work has been partially supported by the 
Grant-in-Aid for Scientific Research (C) No. 11695038 from the Japan 
Society of the Promotion of Science and by Grant-in-Aid of the Russian 
Foundation for Basic Research No. 99-02-18247.

\begin{figure}

\caption{Temperature dependencies of electrical resistivity in 
        ${\rm Ni}_{2+x}{\rm Mn}_{1-x}{\rm Ga}$ ($x = 0 - 0.09$).}

\caption{Temperatures of ferromagnetic ($T_C$), premartensitic ($T_P$) and 
      martensitic ($T_M$) phase transitions as a function of Ni content 
      in the ${\rm Ni}_{2+x}{\rm Mn}_{1-x}{\rm Ga}$ alloys.}

\caption{Features of $\rho(T)$ at martensitic and premartensitic transitions 
     in ${\rm Ni}_{2.04}{\rm Mn}_{0.96}{\rm Ga}$ and 
      ${\rm Ni}_{2.08}{\rm Mn}_{0.92}{\rm Ga}$ upon cooling and heating.}

\caption{Temperature dependence of magnetization $M$ in 
      ${\rm Ni}_{2.02}{\rm Mn}_{0.98}{\rm Ga}$ in the vicinity of the 
      premartensitic transition. The inset shows $M(T)$ in the entire 
     temperature interval.}

\end{figure}

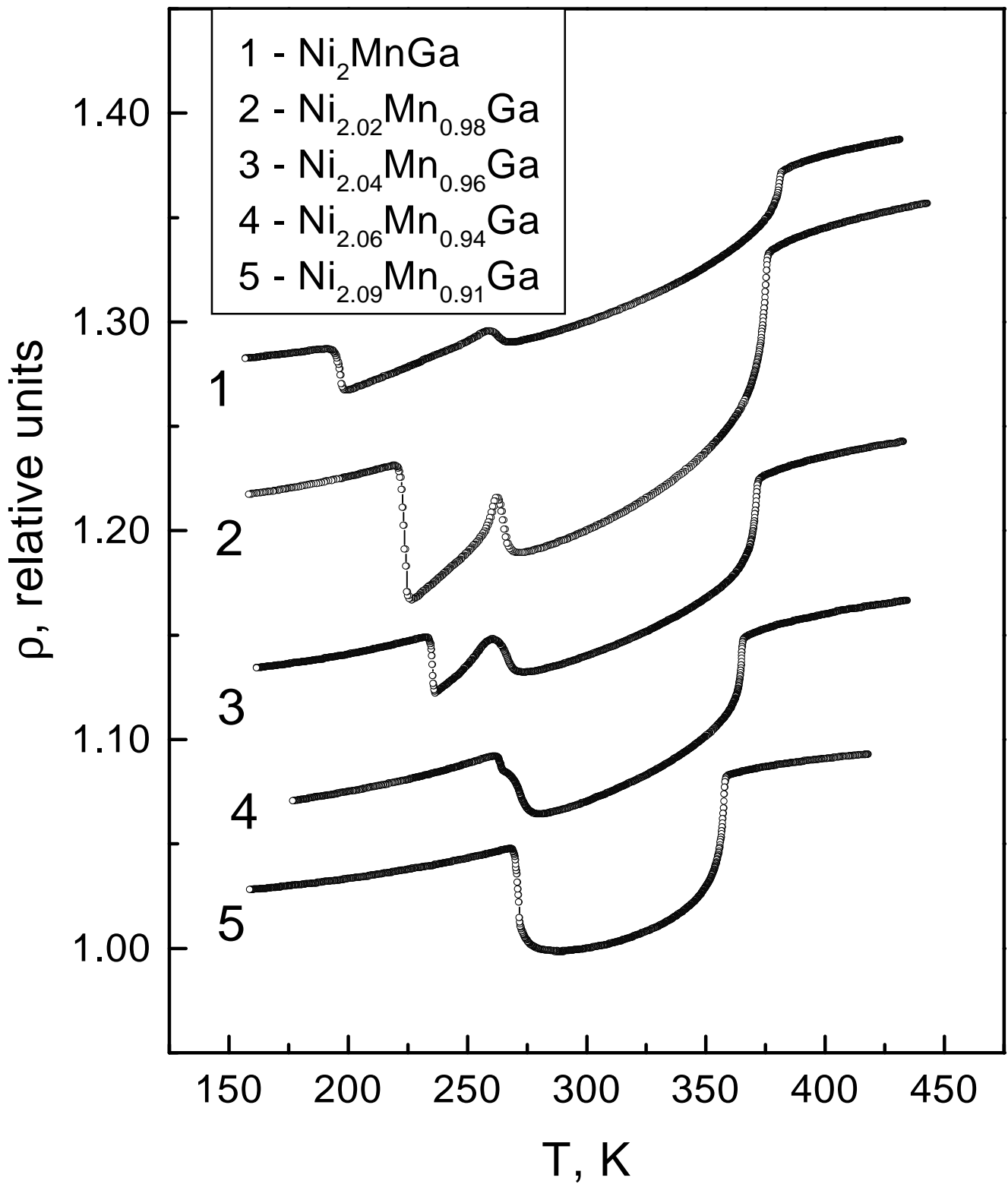
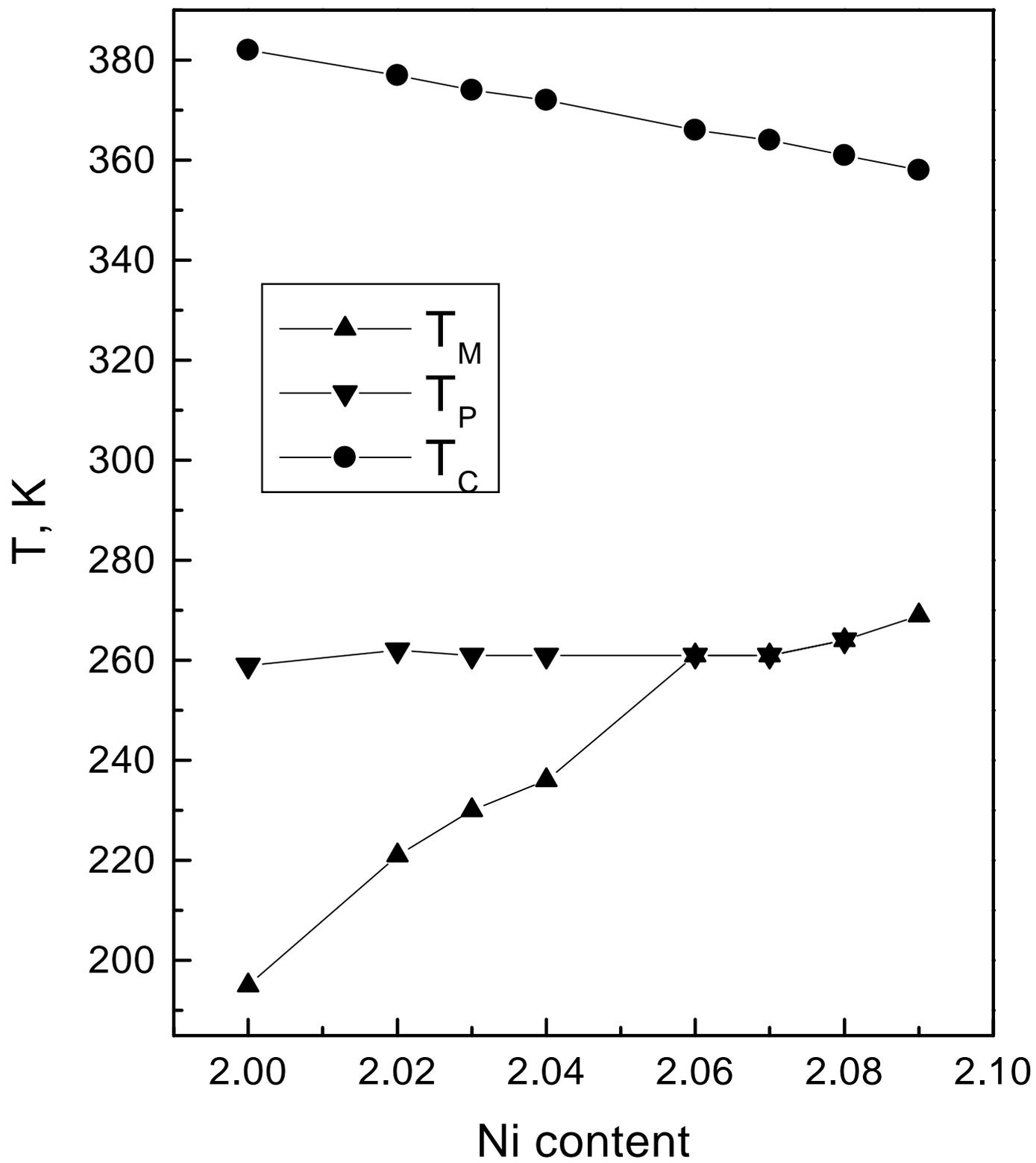
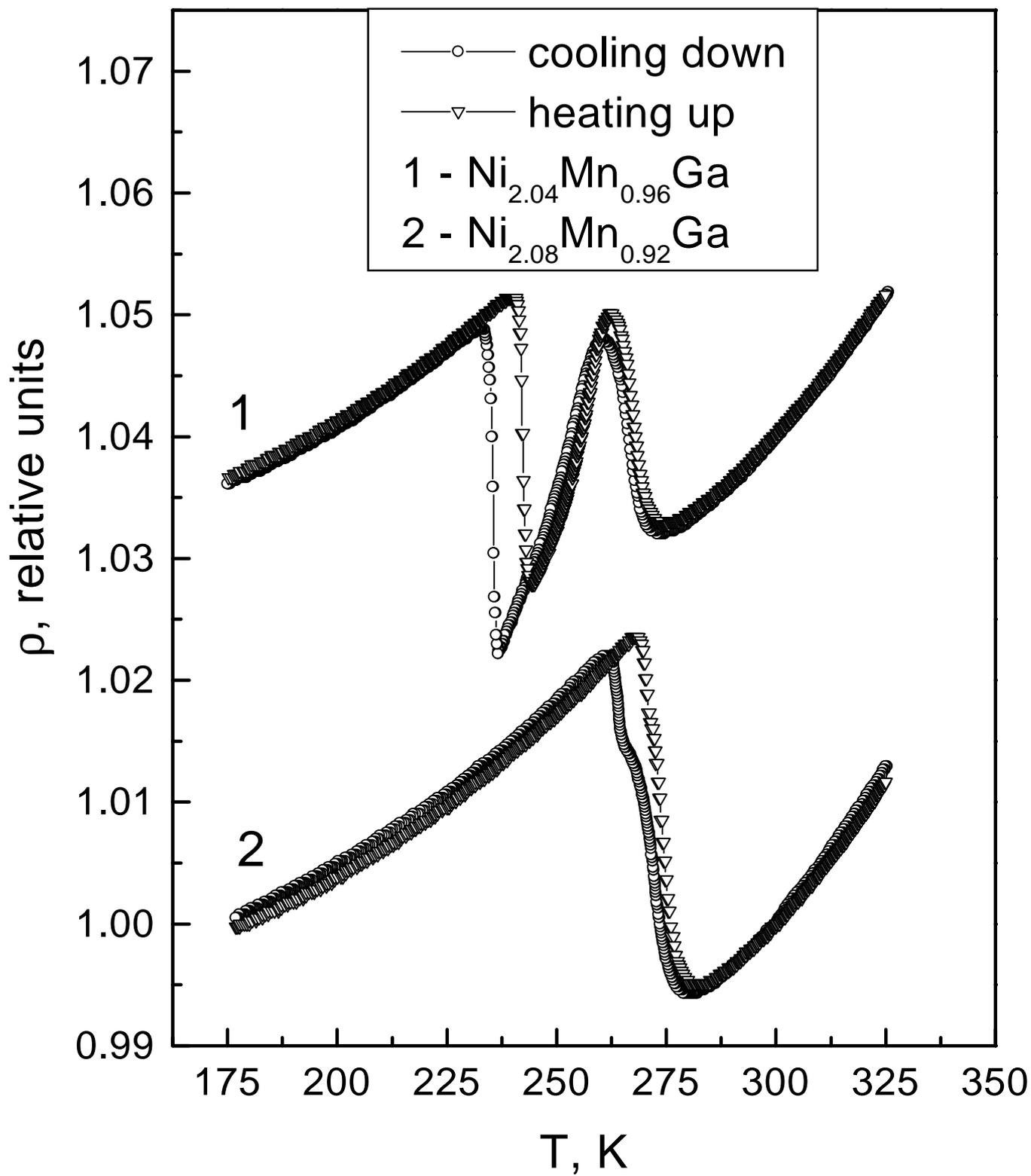
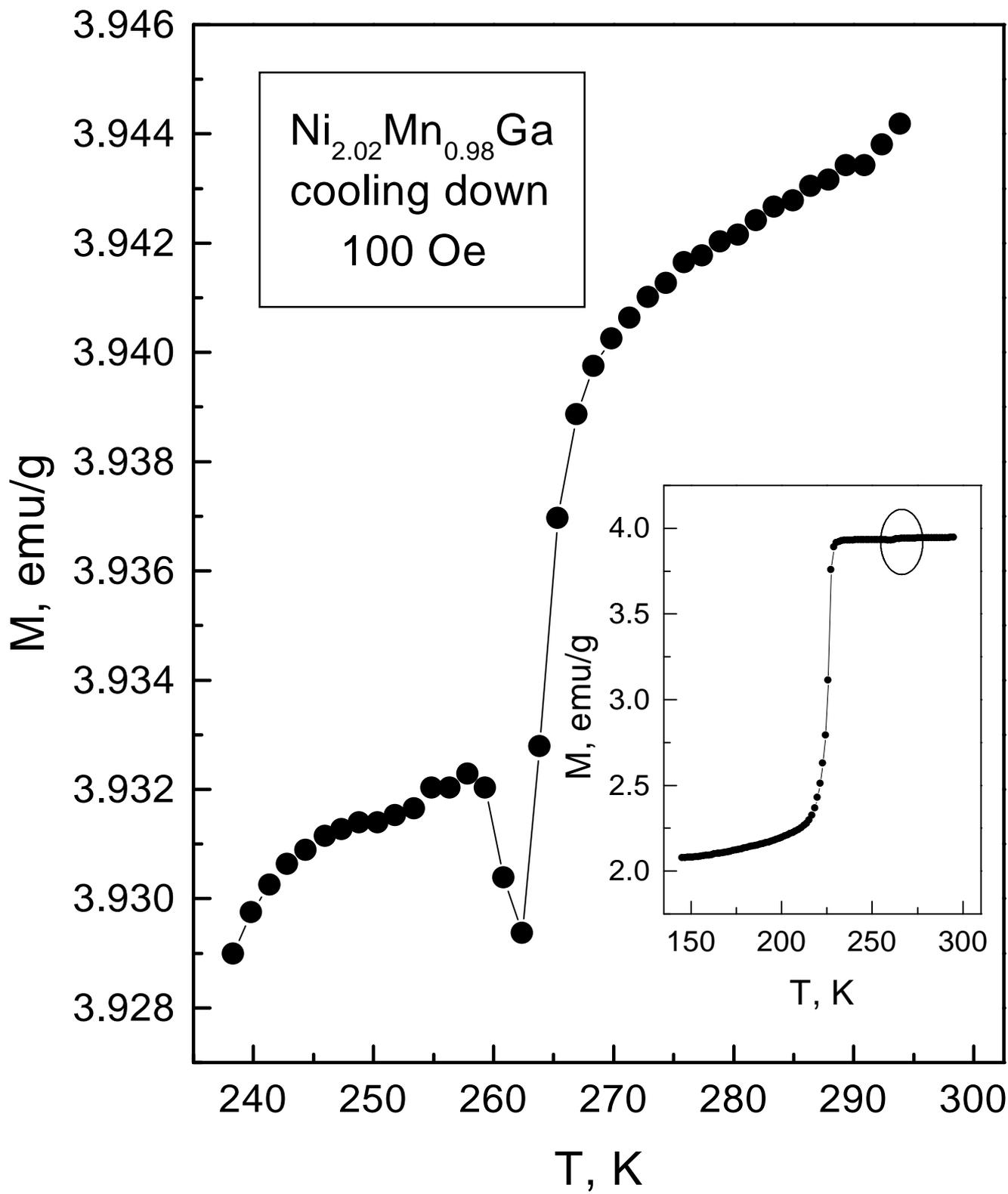
\end{document}